\documentclass[english,manuscript,reprint,showpacs,]{revtex4-1}
\usepackage[T1]{fontenc}
\usepackage[latin9]{inputenc}
\usepackage{color}
\usepackage{babel}
\usepackage{verbatim}
\usepackage{amsmath}
\usepackage{graphicx}
\usepackage{esint}
\usepackage{hyperref}
\usepackage{breakurl}

\makeatletter

\newcommand{\noun}[1]{\textsc{#1}}

\@ifundefined{textcolor}{}
{%
 \definecolor{BLACK}{gray}{0}
 \definecolor{WHITE}{gray}{1}
 \definecolor{RED}{rgb}{1,0,0}
 \definecolor{GREEN}{rgb}{0,1,0}
 \definecolor{BLUE}{rgb}{0,0,1}
 \definecolor{CYAN}{cmyk}{1,0,0,0}
 \definecolor{MAGENTA}{cmyk}{0,1,0,0}
 \definecolor{YELLOW}{cmyk}{0,0,1,0}

\def\Re{\mathrm{Re}}
\def\Im{\mathrm{Im}}

 }

\makeatother

\begin{document}

\preprint{This line only printed with preprint option}

\title{Sub-barrier Coulomb effects on the interference pattern in tunneling ionization photoelectron spectra}

\author{Tian-Min Yan}


\affiliation{Institut f\"ur Physik, Universit\"at Rostock, 18051 Rostock, Germany}
\affiliation{Max-Planck-Institut f\"ur Kernphysik, Postfach 103980, 69029
Heidelberg, Germany}


\author{D.\ Bauer}

\email[Corresponding author: ]{dieter.bauer@uni-rostock.de}

\affiliation{Institut f\"ur Physik, Universit\"at Rostock, 18051 Rostock, Germany}
\begin{abstract}
We use a quantum trajectory-based  semi-classical method to account for Coulomb interaction between the photoelectron and the parent ion in the classically forbidden, sub-barrier region during strong-field tunneling ionization processes.
We show that---besides the well-known modification of the tunneling ionization probability---there is also an influence on the interference pattern in the photoelectron spectra. In the long-wavelength limit, the shift of the intra-cycle interference fringes caused by sub-barrier Coulomb effects in the laser polarization direction can be derived analytically. We compare our results  with \emph{ab initio} solutions of the time-dependent Schr\"odinger equation and find good agreement in the long-wavelength regime, whereas the standard strong field approximation fails. We show that the nodal structure along low-order above-threshold ionization rings is also affected by sub-barrier Coulomb effects.
\end{abstract}


\maketitle

\section{Introduction}

It has been shown experimentally that photoelectron spectra obtained from strong-laser matter interaction contain a wealth of information about the driving laser field and the target \cite{ref_prl_gopal,ref_huisman_science,ref_PRL_perpendicular_carpet,ref_PRL_huismansII}. Decoding this information and discriminating the various spectral features with respect to laser parameters, target geometry, many-electron effects etc.\ is often difficult. On one hand, experiments and \emph{ab initio} solutions of the time-dependent Schr\"odinger equation (TDSE) yield the ``exact'' spectra but little insight. On the other hand, the widely used workhorse of strong-field physics,  the strong field approximation (SFA) \cite{ref_KFR}, allows  for an intuitive interpretation when formulated in terms of interfering quantum orbits via the saddle-point approximation (see, e.g., \cite{ref_ATI_review,ref_JPB_review,mulserbook-chap7,paulusbauerbook}). However, the SFA is rather inaccurate for the case of long-range Coulomb potentials due to the neglect of the interaction between the photoelectron and the parent ion after the time of ionization. In fact, various spectral features originating from this Coulomb interaction have been discovered recently, among them the so-called ``low-energy structure'' \cite{ref_Nature_Phys_LES,ref_PRL_Quan}, asymmetries in elliptically polarized laser fields \cite{ref_Coulomb_asymmetry_circular,ref_Coulomb_asymmetry_circularII}, and holographic side-lobes \cite{ref_huisman_science,ref_PRL_huismansII}.

Within the trajectory-based SFA, the matrix element for ionization is written as a coherent sum over ionization times  for which the corresponding quantum orbits all lead to a given, final momentum $\boldsymbol{p}$  at the detector (see, e.g., \cite{ref_ATI_review,ref_JPB_review,mulserbook-chap7,paulusbauerbook}). If there are more than one such orbits of similar weight, interference comes into play. Interfering orbits originating from subsequent laser cycles give rise to peaks in energy spectra that are separated by $\hbar\omega$ with $\omega$ being the laser frequency, the so-called ``above-threshold ionization'' (ATI) peaks. We call this ``inter-cycle interference'', in contrast to ``intra-cycle interference''  \cite{perel,ref_arbo_inter_intra} where the interfering orbits originate from within one laser cycle, leading to, e.g.,  the holographic interference patterns observed in \cite{ref_huisman_science,ref_PRL_huismansII}. Intra- and inter-cycle interference in time is in analogy to a double slit and a grating in position space, respectively.

Once the SFA is formulated in terms of trajectories, one may Coulomb-correct them \cite{perel,ref_CCSFA_JMO,ref_PRL_TCSFA_LES,ref_huisman_science,ref_Coulomb_asymmetry_circularII}. This is straight-forwardly done in the classically allowed region, i.e., from the so-called ``tunnel exit'' to the detector. In the current paper we will extend the Coulomb-correction of photoelectron spectra to the sub-barrier region where time and orbit are complex. The imaginary time method has since many years been employed for the derivation of Coulomb-corrected tunneling ionization {\em rates} (see \cite{ref-ITM} for a review). In fact, the difference in the probability for  tunneling through a triangular barrier of a short-range potential and through the corresponding Coulomb barrier can be many orders of magnitude \cite{perel2,rate_popru}. However, to the best of our knowledge sub-barrier Coulomb effects on the interference pattern of photoelectron {\em spectra} have not been investigated.

The outline of our paper is as follows. In Section \ref{sec:tcsfa}  we review the trajectory-based Coulomb-corrected strong-field approximation (TCSFA) in real time and space before we introduce, in Subsection \ref{sec:subcc}, the sub-barrier Coulomb correction (sub-CC). In Section  \ref{sec:intracyc} results on the intra-cycle interference are presented. In particular, the fringe shift in polarization direction due to the sub-CC is derived and discussed. There is no sub-CC exactly in perpendicular direction, as discussed in Subsection \ref{sub:perpendicular_direction_phase_analysis}. The nodal structure along low-order ATI rings is investigated in Section \ref{sec:ationring}. We summarize in Section \ref{sec:summ}. Atomic units $\hbar=m_e=|e|=4\pi\epsilon_0=1$ are used, unless noted otherwise.

\section{Trajectory-based Coulomb-corrected strong-field approximation (TCSFA)} \label{sec:tcsfa}
A quantum orbit in plain SFA is a semi-classical trajectory of a photoelectron starting at a complex time $t_{s}$, subsequently governed
by the external light field, and finally detected with the asymptotic
momentum $\boldsymbol{p}$.
 The transition amplitude between
the bound state to the continuum state of asymptotic momentum $\boldsymbol{p}$
is a coherent superposition of contributions from all relevant quantum
orbits (indexed by $\alpha$) that lead to the  asymptotic momentum $\boldsymbol{p}$,
\begin{equation} M_{\boldsymbol{p}}^{(\mathrm{SFA})}\sim\sum_{\alpha}C_{\boldsymbol{p}}(t_{s}^{(\alpha)})\, e^{iW_0(t_{s}^{(\alpha)})}.\label{spsum} \end{equation}
$C_{\boldsymbol{p}}(t_{s}^{(\alpha)})$ is a prefactor whose form is not of interest in this work (see, e.g., \cite{popov_review,ref_JPB_review} for more details). 
\begin{equation}
W_0(t_{s})=-\int_{t_{s}}^{\infty}\left[\frac{1}{2}\boldsymbol{v}^{2}(t)+I_{p}\right]dt 
\end{equation}
is the action in which
\begin{equation}
\boldsymbol{v}(t) = \boldsymbol{p} + \boldsymbol{A}(t)\label{vpA}
 \end{equation}
with the vector potential of the laser field  $\boldsymbol{A}(t)$ in dipole approximation,
and $I_p$ is the ionization potential. 
All the trajectories $\alpha$ leading to an asymptotic electron momentum  $\boldsymbol{p}$ fulfill the equation of motion
\begin{equation} \frac{d\boldsymbol{v}}{dt}=-\boldsymbol{E}(t), \quad \boldsymbol{E}(t) = -\partial_t \boldsymbol{A}(t) \label{eomsfa}\end{equation}
and start at the complex saddle-point (i.e., ionization) time $t_{s}^{(\alpha)}$, which is determined by the saddle point equation (SPE)
\begin{equation}
\left.\frac{\partial}{\partial t}  W_0(t) \right|_{t_s} = 0 \quad \Rightarrow\quad \frac{1}{2}\left[\boldsymbol{p}+\boldsymbol{A}(t_{s}^{(\alpha)})\right]^{2}=-I_{p}.\label{eq:SPE}
\end{equation}
With $I_{p}>0$ and $\boldsymbol{p}$ real-valued, $t_{s}^{(\alpha)}$
is complex. From (\ref{eomsfa}) and (\ref{vpA}) follows that the canonical momentum is conserved, i.e.,
$\boldsymbol{p}=$ const.
The amplitude (\ref{spsum}) clearly demonstrates  the multi-slit in time-nature of quantum mechanical ionization
dynamics \cite{paulusbauerbook} and lays the cornerstone for our further analysis.

The key idea of the TCSFA is to insert the Coulomb force due to the ion of nuclear charge $Z$ into the electronic equation of motion,
\begin{equation} \frac{d\boldsymbol{v}}{dt}=-\boldsymbol{E}(t) - \frac{Z\boldsymbol{r}}{|\boldsymbol{r}|^3}, \label{eomfull}\end{equation}
 and the corresponding Coulomb potential into the action,
\begin{equation}
W(t_{s})=-\int_{t_{s}}^{\infty}\left[\frac{1}{2}\boldsymbol{v}^{2}(t)-\frac{Z}{|\boldsymbol{r}(t)|}+I_{p}\right]dt,\label{eq:action_TCSFA}
\end{equation}
so that the amplitude (\ref{spsum}) becomes
\begin{equation} M_{\boldsymbol{p}}^{(\mathrm{TCSFA})}\sim\sum_{\alpha'}C_{\boldsymbol{p}_0}(t_{s}^{(\alpha')})\, e^{iW(t_{s}^{(\alpha')})}.\label{tcsfaspsum} \end{equation}
Here, we write $\boldsymbol{p}_0$ to indicate that $\boldsymbol{p}\neq$\,const now. Instead $\boldsymbol{p}(t_{s}^{(\alpha')})=\boldsymbol{p}_0$. The saddle-point times are also modified (hence $\alpha'$), as they satisfy
\begin{equation}
\frac{1}{2}\left[\boldsymbol{p}_0+\boldsymbol{A}(t_{s}^{(\alpha')})\right]^{2}=-I_{p} \label{spedash}.
\end{equation}

\begin{figure}
\includegraphics[width=0.45\textwidth]{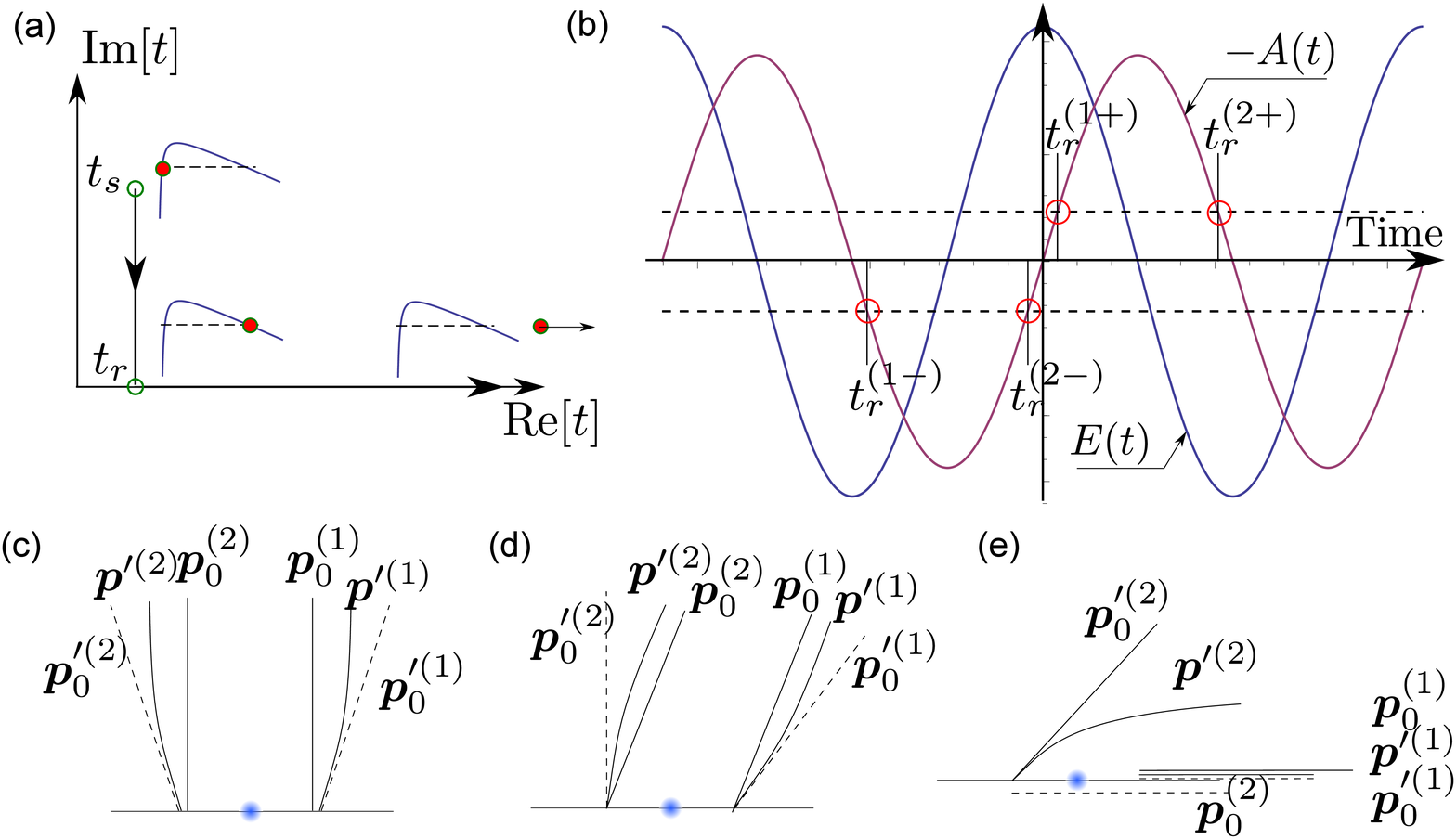}

\caption{(color online). (a) Integration path in the complex time plane. Starting at $t_s$ we integrate parallel to the imaginary time axis down to $t_r$ on the real axis, and then along the real axis to infinity. The corresponding situations in position space around the tunneling barrier are illustrated: at $t_s$ the electron is inside the atom in front of the barrier, at $t_r$ it leaves the barrier at the tunnel exit and subsequently moves in the classically allowed region.  (b) Real parts of the saddle point times for a monochromatic laser
with the electric field $E(t)=E_{0}\cos\omega t$ and vector potential $A(t)=-A_0\sin\omega t$, $E_0=A_0\omega$. The superscripts
of $t_{r}$ indicate  the signs $(+)$, $(-)$ of the
asymptotic momentum $p_z$; the numbers $1$ and
$2$ indicate whether the corresponding quantum trajectory is ``long'' or ``short,'' respectively.   The negative vector potential (red) and the electric field
(blue) in arbitrary units are indicated. Panels (c)--(e) illustrate the asymptotes of real
trajectories in perpendicular direction $\theta=\pi/2$ (c), for an angle $0<\theta<\pi/2$ (d), and parallel to the polarization direction, $\theta=0$ (e). The bold solid lines $\boldsymbol{p}_{0}^{(1)}$ and $\boldsymbol{p}_{0}^{(2)}$ represent trajectories in plain SFA where initial and asymptotic momentum $\boldsymbol{p}$ are equal. Thin solid lines represent the corresponding TCSFA trajectories, which start off like the dashed trajectories with a modified initial momentum but then get Coulomb-distorted such that the  asymptotic momentum is $\boldsymbol{p}$.
\label{fig:demo}}
\end{figure}

 As illustrated in Fig.~\ref{fig:demo}(a),
the integration in complex time in $W(t_s)$ can be splitted into two parts,
\begin{equation} 
W=W^{\mathrm{sub}}+W^{\mathrm{re}}.
\end{equation}
The sub-barrier action $W^{\mathrm{sub}}=-\int_{t_{s}}^{t_{r}} \cdots $, where $t_r =\Re\, t_s $, may be
calculated along the path parallel to the imaginary axis down to the real axis. Provided the integrand is analytic, the result for the total action $W$ is independent of the integration path.  The
action $W^{\mathrm{re}}=-\int_{t_{r}}^{\infty}\cdots $ in the classically allowed region is 
evaluated along the real axis.
 
The well-established imaginary time method (ITM) \cite{ref-ITM} has been used to study quantum tunneling through (time-dependent) barriers in various contexts (e.g., in \cite{ref:hector}).  At $t=t_{s}$, the electron is in the bound state with
dynamical variables in general being complex.
 At $t=t_{r}$, the electron emerges at the outer turning point $\boldsymbol{r}(t_{r})$ (tunnel exit). The boundary conditions for the quantum orbits may be chosen such that from that time $t_r$ on all variables are purely real \cite{realorimaginary}. Let us first summarize the Coulomb-correction to this real part of the propagation, as it has been employed in \cite{ref_CCSFA_JMO,ref_PRL_TCSFA_LES,ref_huisman_science,ref_Coulomb_asymmetry_circularII}.

\subsection{Real-part Coulomb correction}\label{sec:repart}
For $W^{\mathrm{re}}$, the real trajectories with the Coulomb interaction taken into account are obtained by solving the equation of motion (\ref{eomfull})
with the initial conditions 
\begin{equation} 
\boldsymbol{r}(t_{r}) = \boldsymbol{\alpha}(t_r) - \Re\, \boldsymbol{\alpha}(t_s)
\end{equation} (tunnel exit) where
\begin{equation}  \boldsymbol{\alpha}(t) = \int^t \boldsymbol{A}(t')\, dt' \end{equation}
 is the electron excursion, and 
\begin{equation} 
\boldsymbol{v}(t_{r}) = \boldsymbol{p}_0 + \boldsymbol{A}(t_r). \end{equation}
 The corresponding saddle-point times $t_{s}^{(\alpha')}$ are determined via Eq.~(\ref{spedash}). The equations of motion are solved numerically, along with $W^{\mathrm{re},(\alpha')}$, up to the end of the laser pulse. The asymptotic momentum $\boldsymbol{p}^{(\alpha')}$ can then be determined via Kepler's formulas. The values of $\boldsymbol{p}^{(\alpha')}$, $e^{i W^{\mathrm{re},(\alpha')}}$, and other entities of interest are saved in a data base. In the previous works \cite{ref_PRL_TCSFA_LES,ref_huisman_science,ref_Coulomb_asymmetry_circularII} the uncorrected $ W^{\mathrm{sub},(\alpha')}$ (i.e., without Coulomb potential) of the plain SFA with $\boldsymbol{p}=\boldsymbol{p}_0$ was evaluated analytically. Finally, photoelectron spectra $| M_{\boldsymbol{p}}|^2$ can be calculated by adding coherently all the $M_{\boldsymbol{p}}^{(\alpha')}$ of the quantum orbits from the data base that have their asymptotic momentum $\boldsymbol{p}^{(\alpha')}$ in a bin around the final momentum of interest $\boldsymbol{p}$. The size of the bin determines the resolution of the photoelectron momentum spectrum. 
In the current work we used $1600\times400$ bins to sample the final momentum interval $[-2,2]\times[0,1]$ in the spectra of Fig.~\ref{fig:spec_2D_sin2_2um_1e14_nc4}(b)--(d) below. Thanks to the azimuthal symmetry with respect to the laser polarization direction $\boldsymbol{e}_z$ it is sufficient to consider only, e.g., $p_{x}\geq 0$.
We covered randomly the initial momentum interval $[-2,2]\times[0,2]$ by 32 million points. For each initial momentum several saddle-points contribute. In the case of the 3-cycle pulse considered below we thus dealt with $\simeq 100$ million trajectories.  Because of focusing along the polarization axis the randomly sampled initial momenta in lateral direction $p_{x0}$ should cover a larger interval than the final $p_x$ of interest. On the other hand,  it rarely occurs that $|p_{z0}|>|p_z|$ so that the initial  momentum interval in polarization direction may be chosen the same as the final one.  If, for an initial momentum $p_{x0}>0$ the final momentum $p_x$ turns out to be $< 0$ we simply reverse the signs of both $p_{x0}$ and $p_x$. The finer the binning and the higher the dynamic range of interest the more trajectories are required to ensure good enough statistics. It is also advisable to shoot randomly in the initial momentum plane, not in a regular, grid-like pattern, in order to avoid artificial structures in the final momentum spectra.
The numerical effort to obtain TCSFA spectra like the one in Fig.~\ref{fig:spec_2D_sin2_2um_1e14_nc4}(c) by solving the equations of motion for 100 million trajectories is significant but trivially parallelizable \cite{linuxcluster}.

\subsection{Sub-barrier Coulomb correction}\label{sec:subcc}
One may attempt to solve the equation of motion (\ref{eomfull}) in complex space and time under the barrier as well, for instance, by propagating from the tunnel exit ``backwards'' in imaginary time from $t_r$ to $t_s$. However, we found that the photoelectron spectra obtained in this manner are completely spoiled by artifacts such as caustics, destroying the interference pattern observed in the {\em ab initio} TDSE results. If one wanted to propagate in forward direction from $t_s$ to $t_r$ it is not clear what the initial conditions should be. Choosing, e.g., $\Re\, r(t_s)=0$ one would have to iterate $\Im\, \boldsymbol{r}(t_s)$ and the imaginary $\boldsymbol{v}(t_s)$ such that at the tunnel exit position and velocity become real, and $\boldsymbol{p}_0=\boldsymbol{v}(t_r) - \boldsymbol{A}(t_r)$ because $t_s$ is determined by $\boldsymbol{p}_0$. In the derivation of tunneling ionization {\em rates} \cite{perel2,ref-ITM,rate_popru}  such problems are circumvented by matching the action  $W^{\mathrm{sub}}$,  evaluated with the zeroth order (i.e., plain SFA) quantum orbit with the highest statistical weight, to the exponent of the unperturbed, initial state wavefunction. This is justified, as for $r\to 0$ the Coulomb potential certainly dominates the external laser field. Moreover, the quantum orbits are almost linear \cite{ref_bondar_sub_traj} and spatially
confined in the polarization direction, pointing from the origin to
the tunnel exit. We pursue a similar strategy in this work in so far that we consider the plain-SFA quantum orbits for the evaluation of  $W^{\mathrm{sub}}$ only. Of course, for the calculation of photoelectron {\em spectra} we cannot restrict ourselves to just the dominating quantum orbits.

The  plain-SFA quantum orbits fulfill
\begin{equation} 
\boldsymbol{v}(t)=\boldsymbol{p}_{0}+\boldsymbol{A}(t)
\end{equation}
where $\boldsymbol{p}_{0}$ is the randomly sampled momentum from the real-part propagation described in the previous Subsection.
 The sub-barrier trajectory is 
\begin{equation}
\boldsymbol{r}(t)=\int_{t_{s}}^{t}\boldsymbol{v}(t')dt'-i\,\Im\int_{t_{s}}^{t_{r}}\boldsymbol{v}(t')dt',\label{eq:complex_traj}
\end{equation}
which ensures that $\boldsymbol{r}(t_r)$ is purely real and $\boldsymbol{r}(t_s)$ is purely imaginary.
We write the sub-barrier action as
\begin{equation} W^{\mathrm{sub}}=W^{\mathrm{sub,0}}+W^{\mathrm{subcc}},\end{equation}
where
\begin{equation}
W^{\mathrm{sub,0}}=-\int_{t_{s}}^{t_{r}}\left[\frac{1}{2}v^{2}(t)+I_{p}\right]dt\label{eq:w_sub_0}
\end{equation}
is the plain-SFA action, and
\begin{equation}
W^{\mathrm{subcc}}=\int_{t_{s}}^{t_{r}}\frac{Z}{\sqrt{\boldsymbol{r}^{2}(t)}}\,dt\label{eq:w_subcc}
\end{equation}
is the sub-CC action. Here, we write on purpose $Z/\sqrt{\boldsymbol{r}^{2}(t)}$ instead of the non-analytic $Z/|\boldsymbol{r}(t)|$.

In the preceding studies \cite{ref_PRL_TCSFA_LES,ref_huisman_science,ref_Coulomb_asymmetry_circularII}   only $W=W^{\mathrm{sub},0}+W^{\mathrm{re}}$ was taken into account whereas in  this work, the action
takes the full form
\begin{equation}
W=W^{\mathrm{sub},0}+W^{\mathrm{subcc}}+W^{\mathrm{re}},\label{eq:action_w}
\end{equation}
including the sub-CC action $W^{\mathrm{subcc}}$.
 
The action may diverge if the complex trajectory $\boldsymbol{r}(t)=\boldsymbol{0}$. However, thanks  to the negligible probability
of sampling momenta $\boldsymbol{p}_0$ that lead to such singularities they turned out to be harmless in the actual numerical implementation of our sub-CC.

\section{Intra-cycle interference} \label{sec:intracyc}
In order to investigate the influence of the sub-CC on the interference pattern in photoelectron momentum spectra  we first consider H(1s) ($I_{p}=0.5$) in a linearly polarized laser pulse.
The vector potential of the laser pulse of carrier frequency  $\omega$ is given  (in dipole approximation) by 
\begin{equation}
A(t)=-A_0\,\sin^{2}\left(\frac{\omega t}{2N_{c}}\right)\sin\omega t,
\end{equation}
for $0<t<N_{c} \, 2\pi/\omega$ and zero otherwise.
$E_{0}=A_0\omega $ is the peak electric
field strength, $N_{c}$ is the number of laser cycles.

\begin{figure}
\includegraphics[scale=0.38]{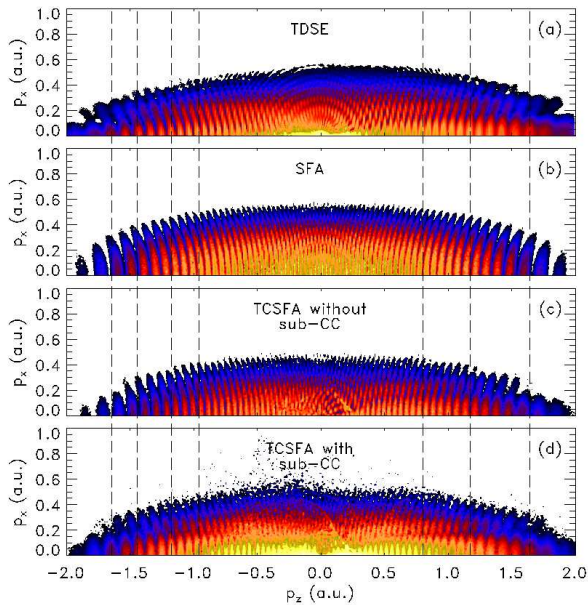}

\caption{(color online). Logarithmically scaled photo-electron momentum
distribution in the $p_{z}$-$p_{x}$ plane ($p_{z}\equiv p_{\parallel}$
along the polarization direction) calculated using (a) TDSE, (b) plain SFA,
(c) TCSFA without sub-CC, and (d) TCSFA with sub-CC. The calculation
is for H(1s) in a 4-cycle, $\sin^{2}$-envelope, near-infrared
laser pulse of wavelength $\lambda=2\,\mu$m and intensity $10^{14}$
W/cm$^{2}$. Four orders of magnitude in probability are shown. \label{fig:spec_2D_sin2_2um_1e14_nc4}}
\end{figure}

Figure~\ref{fig:spec_2D_sin2_2um_1e14_nc4}(a)
shows the photoelectron momentum spectrum, calculated from the exact numerical solution of  the time-dependent Schr\"{o}dinger equation
(TDSE) using the \noun{Qprop} code \cite{ref_qprop}. The laser parameters are $N_{c}=4$, $\omega=0.0228$, $E_{0}=0.0534$ (corresponding to $\lambda=2$ $\mu$m and an intensity of $10^{14}$\,W/cm$^{2}$). Here we focus on the almost vertically aligned interference fringes \cite{remark:sidelobes}.
 These fringes stem from the ``intra-cycle interference'' \cite{perel,ref_arbo_inter_intra} of two trajectories ``born'' at times $t_{r}^{(1)}$
and $t_{r}^{(2)}$ at the respective tunnel exits within one laser cycle. The electric field has opposite signs at $t_{r}^{(1)}$  and $t_{r}^{(2)}$, meaning that the tunnel exits point in opposite directions. In the context of
the SFA, the two trajectories are typically denoted by ``short'' and ``long''
according to whether the signs of the projections of $\boldsymbol{r}(t_{r})$ and
$\boldsymbol{p}_{0}$ onto the polarization direction are equal or opposite, respectively. In the former case the photoelectron drifts directly from the tunnel exit to the detector whereas in the latter case it starts off into the ``wrong'' direction before it is reversed by the laser field.

The vertical fringes are also present in the SFA spectrum  Fig.~\ref{fig:spec_2D_sin2_2um_1e14_nc4}(b). However, the positions are shifted  relative to the TDSE result, as indicated by the vertical, dashed lines. In fact, when there is constructive interference in the TDSE result, there is destructive interference in the SFA spectrum, and {\em vice versa}. This indicates that there is an erroneous phase-shift  between the long trajectory and the short trajectory of $\pi$ in the plain SFA as compared to the exact TDSE result. It is expected that incorporating the Coulomb-attraction between the outgoing electron and the parent ion should cure this disagreement. 

Let us now ``switch-on'' the Coulomb-attraction for the propagation in the classically allowed region (i.e., from the tunnel exit to the detector), as described in Section \ref{sec:repart}. The result is shown in Fig.~\ref{fig:spec_2D_sin2_2um_1e14_nc4}(c). As discovered previously \cite{ref_huisman_science}, side-lobes are reproduced in that way. However, the vertical interference fringes still have the same position as in the plain-SFA result, meaning that the erroneous phase-shift  between the long and the short trajectory of $\pi$ is {\em not} cured by the Coulomb-correction out of the Coulomb-barrier. Only if the sub-CC, as described in Section \ref{sec:subcc}, is incorporated as well, do the fringe positions coincide  with the {\em ab initio} TDSE result [see Fig.~\ref{fig:spec_2D_sin2_2um_1e14_nc4}(d)]. Hence, the sub-CC not only affects the statistical weight of a quantum orbit via the imaginary part of $W^{\mathrm{subcc}}$ but also the interference pattern in photoelectron spectra via the real part of $W^{\mathrm{subcc}}$. This finding (and its analytical derivation in the following) is the main result of this paper.

Let us suppose that only two quantum orbits are dominating the transition amplitude for a certain asymptotic momentum $\boldsymbol{p}$. Given the two individual transition amplitudes $M_{\boldsymbol{p}}^{(1)}$ and $M_{\boldsymbol{p}}^{(2)}$, the probability to find a photoelectron with asymptotic momentum $\boldsymbol{p}$ is $w(\boldsymbol{p})=\left[M_{\boldsymbol{p}}^{(1)}+M_{\boldsymbol{p}}^{(2)}\right]^{*}\left[M_{\boldsymbol{p}}^{(1)}+M_{\boldsymbol{p}}^{(2)}\right]$
with $M_{\boldsymbol{p}}^{(\alpha')}\sim e^{iW^{(\alpha')}}(\alpha'=1,2)$.
The action $W^{(\alpha')}$ is of the form (\ref{eq:action_w}).
Interference with pronounced contrast
requires the weights (i.e., absolute values) of $M_{\boldsymbol{p}}^{(1)}$ and $M_{\boldsymbol{p}}^{(2)}$
being comparable, $e^{-\mathrm{Im}W^{(1)}}\sim e^{-\mathrm{Im}W^{(2)}}$. The assumption that this is the case then 
yields  an ionization probability $w(\boldsymbol{p})\simeq 2e^{-2\mathrm{Im}W^{(1)}}(1+\cos\phi)$. Hence,  the interference pattern is determined by the phase difference $\phi=\mathrm{Re}[W^{(2)}-W^{(1)}]$. 
For $\phi=m\pi$ with $m$ even we have interference maxima, for $m$ odd interference minima, provided the prefactors $C_{\boldsymbol{p}}(t_{s}^{(\alpha)})$ in (\ref{spsum}) do not introduce extra phase-differences \cite{remark:prefactorphases}.

The separation of the action (\ref{eq:action_w}) suggests that  we split the phase  $\phi$ accordingly,
\begin{equation}
\phi=\phi^{\mathrm{sub,0}}+\phi^{\mathrm{subcc}}+\phi^{\mathrm{re}}.\label{eq:phase_angle}
\end{equation}
From the simulation results in Fig.~\ref{fig:spec_2D_sin2_2um_1e14_nc4} we know already that it is $\phi^{\mathrm{subcc}}$ that is responsible for the intra-cycle fringe shifts towards the correct positions, as predicted by the {\em ab initio} TDSE result.  In the following we will derive an analytical expression for  $\phi^{\mathrm{subcc}}$ in the long-wavelength limit.

\subsection{Fringe shift in polarization direction}
\label{sub:polarization_direction_phase_analysis}
  In the
long-wavelength limit ($\omega\rightarrow0$) it is possible to evaluate the sub-CC phase difference $\phi^{\mathrm{subcc}}$
in the polarization direction analytically. Consider a monochromatic laser field
\begin{equation}
E(t)=E_{0}\cos \omega t, \quad A(t)=-A_0 \sin\omega t \label{eq:mono_pulse}
\end{equation}
where $E_{0}=A_{0}\omega$, and we assume $A_0 > 0$. We will prove in this Subsection that in the long-wavelength limit we obtain a momentum-independent sub-CC phase shift of
\begin{equation}
|\phi^{\mathrm{subcc}}|=\frac {Z\pi}{\sqrt{2I_{p}}} . \label{eq:claim}
\end{equation}
In particular, for the case of Fig.~\ref{fig:spec_2D_sin2_2um_1e14_nc4} where $Z=2I_p=1$ we have $|\phi^{\mathrm{subcc}}|=\pi$, turning constructive interference maxima of the plain SFA into destructive interference minima in the TCSFA with sub-CC and {\em vice versa}. 
As we know already from the results in Fig.~\ref{fig:spec_2D_sin2_2um_1e14_nc4} that the Coulomb correction of the propagation in real space is {\em not} responsible for the fringe shift, we ignore its effect on $\boldsymbol{p}_{0}^{(1)}$ and $\boldsymbol{p}_{0}^{(2)}$ (and thus also on $t_{s}^{(1)}$ and $t_{s}^{(2)}$) in this Section and assume \cite{remark:realpartunimportant}
\begin{equation}
\boldsymbol{p}_{0}^{(1)}\simeq \boldsymbol{p}_{0}^{(2)}=\boldsymbol{p}.\label{eq:approxp1p2p}
\end{equation}
For a monochromatic laser field (\ref{eq:mono_pulse}), $\boldsymbol{p}=(p_{x},p_{z })$, and $t_{s}=t_{r}+it_{i}$  the saddle point equation (\ref{eq:SPE}) yields 
\begin{eqnarray}
\sin\left(\omega t_{r}\right)\cosh\left(\omega t_{i}\right) & = & \frac{p_{z}}{A_{0}},\label{eq:SPE_reduced1}\\
\cos\left(\omega t_{r}\right)\sinh\left(\omega t_{i}\right) & =& \mp\frac{\sqrt{2I_{p}+p_{x}^{2}}}{A_{0}}.
\label{eq:SPE_reduced2}
\end{eqnarray}
Equation~(\ref{eq:SPE_reduced1}) shows that  the possible $t_{r}$ can be determined as  the intersections between $p_{z}/\cosh(\omega t_{i})\stackrel{\omega\to 0}{\simeq} p_z$ and the negative vector potential $-A(t)=A_0 \sin\omega t$, as indicated  in Fig.~\ref{fig:demo}(b). The figure shows that a pair of saddle points per cycle,
e.g., $t_{s}^{(1+)}$ and $t_{s}^{(2+)}$, exists for a specified $p_z$ (dashed horizontal lines). The so-called intra-cycle interference is thus due to two  saddle-points for which 
\begin{equation}
\begin{aligned}\boldsymbol{p}^{(1)} & =\boldsymbol{p}^{(2)}=\boldsymbol{p},\\
t_{r}^{(1)} & =\frac{\pi}{\omega}-t_{r}^{(2)},\\
t_{i}^{(1)} & =t_{i}^{(2)}=t_{i}
\end{aligned}
\label{eq:SFA_condition}
\end{equation}
are fulfilled.
In the polarization direction, $p_{x}=0$, the sub-CC action  is 
\begin{eqnarray}
W^{\mathrm{subcc}}=\int_{t_{s}}^{t_{r}}\frac{Z}{\sqrt{z^2(t)}}\, dt=i\int_{t_{i}}^{0}\frac{Z}{\sqrt{\tilde{z}^2(\tau)}}\,d\tau \label{eq:wccsub}
\end{eqnarray}
with  $\tilde{z}(\tau) = z(t_{r}+i\tau)$ along the integration path $t=t_{r}+i\tau,\,\tau\in\left[t_{i},0\right]$,
as shown in Fig. \ref{fig:demo}(a). The superscript $(\alpha)$
is suppressed here. 
For the sub-barrier trajectory in polarization direction calculated from (\ref{eq:complex_traj})
\begin{equation}
z(t)=\frac{A_{0}}{\omega}\left[\cos\omega t-\mathrm{Re}\,\cos\omega t_{s}\right]+p_{z}\left[t-\mathrm{Re}\,t_{s}\right]\label{eq:complex_traj_mono_z}
\end{equation}
results.
Separating the real and imaginary parts
of $\tilde{z}(\tau)$, 
\begin{eqnarray}\tilde{z}(\tau)=a(\tau)+ib(\tau),\label{eq:aandb}\end{eqnarray} 
one finds
\begin{equation}
\begin{aligned}a(\tau) &=  \frac{A_{0}}{\omega}\cos\omega t_{r}\left(\cosh\omega\tau-\cosh\omega t_{i}\right),\\
b(\tau) & =  p_{z}\tau-\frac{A_{0}}{\omega}\sin\omega t_{r}\sinh\omega\tau.
\end{aligned}
\label{eq:A_and_B}
\end{equation}
In the long wavelength limit $\omega\rightarrow0$, the functions
$\sinh$ and $\cosh$ in $a(\tau)$ and $b(\tau)$ can be expanded about $\omega=0$, leading to
\begin{equation}
\begin{aligned}a(\tau) & \simeq  \frac{A_{0}}{2}\omega\cos\omega t_{r}\left(\tau^{2}-t_{i}^{2}\right),\\
b(\tau) & \simeq  \tau\Delta
\end{aligned}
\label{eq:A_and_B_in_limit}
\end{equation}
with
\begin{equation} \Delta\simeq p_{z}-A_{0}\sin\omega t_{r} \simeq \frac{A_0}{2} \omega^2 t_i^2 \sin\omega t_r.\label{eq:Delta} \end{equation}
Let us consider emission times $t_r$ for positive $p_z$. In this case  $0 < \omega t_r < \pi$ and thus $\Delta > 0$. As $\tau >0 $ \cite{ref:llqm} as well, $b(\tau) > 0$. Hence  the integrand
in $W^{\mathrm{subcc}}$ in (\ref{eq:wccsub}) may be simplified [by choosing the positive root] to $Z\left[\tilde{z}^2(\tau)\right]^{-1/2}=Z/\tilde{z}(\tau)$
for $a(\tau)\ge0$ and $Z\left[\tilde{z}^2(\tau)\right]^{-1/2}=-Z/\tilde{z}(\tau)$
for $a(\tau)<0$ so that it is sufficient to consider the simplified action
\begin{eqnarray}
W_{\pm}^{\mathrm{subcc}'}=\pm i\int_{t_{i}}^{0}\frac{Z}{\tilde{z}(\tau)}\,d\tau .
\end{eqnarray}
The real part of this action,
\begin{equation}
\mathrm{Re}W_{\pm}^{\mathrm{subcc}'}=\pm\int_{t_{i}}^{0}\frac{ Z\,  b(\tau)}{a^{2}(\tau)+b^{2}(\tau)}\, d\tau,\label{eq:action_W_polarization_simplified}
\end{equation}
is relevant for interference patterns in the spectra. Inserting (\ref{eq:A_and_B_in_limit}) leads with (\ref{eq:SPE_reduced2}) for $p_x=0$ and $\omega\to 0$,
\begin{eqnarray}
A_{0}\omega t_{i}\cos\omega t_{r}\simeq\mp\sqrt{2I_{p}},\label{eq:2Ip}
\end{eqnarray}
to
\begin{widetext}
\begin{equation}
\mathrm{Re}W_{\pm}^{\mathrm{subcc}'}  =  \pm\frac{2Zt_{i}^{2}\Delta}{I_{p}}\int_{t_{i}}^{0}\frac{\tau\,d\tau}{(\tau^{2} - t_i^2)^2 + 2\frac{\Delta^{2}}{I_{p}}t_i^2\tau^2} \\
=  \left.\pm \frac{Z}{\sqrt{2I_p-\Delta^2}} \,\arctan \left[\frac{I_p \tau^2 - (I_p-\Delta^2)t_i^2}{\sqrt{2I_p-\Delta^2} \,\Delta t_i^2} \right]\right|_{t_i}^0\\
\stackrel{\Delta\to 0}{\simeq }  \mp \frac{Z}{\sqrt{2I_p}} \, \frac{\pi}{2}.
\label{eq:ReWsubccpm}
\end{equation}
\end{widetext}
It follows from (\ref{eq:A_and_B_in_limit}) and (\ref{eq:SFA_condition}) that, for a given $p_z$, the two saddle-point times $t_{s}^{(1)}$ and $t_{s}^{(2)}$ are such that  $a^{(1)}(\tau)=-a^{(2)}(\tau)$ and $b^{(1)}(\tau)=b^{(2)}(\tau)$. The opposite signs of $a(\tau)$ have been considered already in our calculation of $\Re\, W_{\pm}^{\mathrm{subcc}'}$. Hence
\begin{equation}
\left|\phi^{\mathrm{subcc}}\right|=\left|\Re\,W_{+}^{\mathrm{subcc}'} - \Re\,W_{-}^{\mathrm{subcc}'} \right| \simeq \frac{Z \pi}{\sqrt{2I_{p}}} ,
\end{equation}
as claimed in (\ref{eq:claim}).

For a hydrogen-like ion in a state with principal quantum number $n=1,2,3, \ldots$ we have $I_p=Z^2/(2n^2)$ and thus $\left|\phi^{\mathrm{subcc}}\right|=n\pi$. In particular, for ground states $n=1$ we have independently of $Z$ that  $\left|\phi^{\mathrm{subcc}}\right|=\pi$. This is the required phase shift for bringing the intra-cycle fringe pattern in agreement with the {\em ab initio} TDSE result [see Fig.~\ref{fig:spec_2D_sin2_2um_1e14_nc4}(d)].

We compared TDSE and TCSFA results for other parameters $Z$, $I_p$, and $n$ and find good agreement if the sub-CC is included, provided the minimum number of photons required for ionization $I_p/\omega\gg 1$ and the Keldysh parameter $\sqrt{I_p/(2U_p)}\lesssim 1$, where $U_p=A_0^2/4$ is the ponderomotive energy. Moreover, the smaller $Z$ the better is the approximation to use the zeroth-order, plain-SFA trajectory for the sub-CC. Excited states $n>1$ may introduce extra phase-differences via the prefactors $C_{\boldsymbol{p}}(t_{s}^{(\alpha)})$ in (\ref{spsum}) \cite{remark:prefactorphases}, not discussed in this work.

The reader may have noticed that the precursors of ATI rings, visible in  the exact TDSE result of Fig.~\ref{fig:spec_2D_sin2_2um_1e14_nc4}(a) and in the SFA result (b) as circular structures around the origin, are actually less clear in the TCSFA (c), let alone in the sub-CC TCSFA (d). The reasons for this deterioration are the overemphasized classical boundaries and caustics that spoil for few-cycle pulses the low-energy part of the spectra by rendering the few trajectories that contribute to the ATI-ring structure relatively less important. The situation improves for more-cycle pulses, as in Fig.~\ref{fig:spec_2D_const_800nm_1e14} below, where more trajectories contribute to inter-cycle features such as ATI rings.

\subsection{Interference in perpendicular direction}\label{sub:perpendicular_direction_phase_analysis}
We are also interested
in effects from the sub-CC in the perpendicular direction $p_{z}=0$.
In the SFA, the short and long trajectories are asymptotically parallel, as shown
in Figs.~\ref{fig:demo}(c--e). In the perpendicular direction [Fig.~\ref{fig:demo}(c)], the parallel asymptotes are vertical when
two saddle points merge as $p_{z}\rightarrow0$ in (\ref{eq:SPE_reduced1}). One  easily verifies that in Fig.~\ref{fig:demo}(b) $t_r^{(1+)}$ merges with $t_r^{(2-)}$ and   $t_r^{(2+)}$ merges with $t_r^{(1-)}$. Hence, in plain SFA the momenta and the saddle point times of long and short trajectories become degenerate in perpendicular direction [$\boldsymbol{p}_{0}^{(1)}$  and $\boldsymbol{p}_{0}^{(2)}$ in  Fig.~\ref{fig:demo}(c)]. However, with the long-range Coulomb interaction included, this degeneracy is lifted. In order to have a common asymptotic momentum ${\boldsymbol{p}}'$ the initial momenta ${\boldsymbol{p}}'_0$ must be different, as indicated in    Fig.~\ref{fig:demo}(c).

In order to cancel the longitudinal shift induced by the long-range
Coulomb distortion, the two trajectories should satisfy 
\begin{equation}
p_{0z}^{(1)}=-p_{0z}^{(2)},\qquad p_{0x}^{(1)}=p_{0x}^{(2)}.
\end{equation} 
The corresponding
saddle points are easily found for these criteria from Fig.~\ref{fig:demo}(b). Taking, e.g., the saddle
point $(1+)$ as the reference, the other saddle point is $(1-)$.
Therefore, the initial conditions for these trajectories are
\begin{equation}
\begin{aligned}p_{0z}^{(1+)}=-p_{0z}^{(1-)}, & \qquad p_{0x}^{(1+)}=p_{0x}^{(1-)},\\
t_{r}^{(1+)}=\frac{\pi}{\omega}+t_{r}^{(1-)}, & \qquad t_{i}^{(1-)}= t_{i}^{(1+)}.
\end{aligned}
\label{eq:perpendicular_conditions}
\end{equation}

In the following, the phase contributions $\phi_{\mathrm{c}\perp}^{\mathrm{subcc}}$,
$\phi_{\mathrm{c}\perp}^{\mathrm{sub},0}$ and $\phi_{\mathrm{c}\perp}^{\mathrm{re}}$
are deduced, and the interference pattern in the perpendicular
direction is inferred. The subscript ``$\mathrm{c}$'' indicates
that the phase difference between the  trajectories has been generalized
to allow for the real-space trajectory-distortion by the Coulomb force.
The interference pattern in perpendicular direction
is determined by the symmetry properties of the interfering trajectories, while it is
 independent of the detailed, Coulomb-modified dynamics because both trajectories of a pair collect equal phase distortions. This suggests that the
SFA, TCSFA with or without sub-CC result in the same interference structure
in the perpendicular direction, as will be discussed in the following.

\subsubsection{Phase difference from sub-CC: $\phi_{\mathrm{c}\perp}^{\mathrm{subcc}}$}\label{sub:perp_sub_CC}
We show first that the sub-CC phase difference $\phi_{\mathrm{c}\perp}^{\mathrm{subcc}}$ does not contribute in (\ref{eq:phase_angle}),
\begin{equation}
\phi_{\mathrm{c}\perp}^{\mathrm{subcc}}=\mathrm{Re}\left[W_{\perp}^{\mathrm{subcc}(1+)}-W_{\perp}^{\mathrm{subcc}(1-)}\right]=0.\label{eq:phi_c_subcc_perpendicular}
\end{equation}
The real part of the sub-CC action reads 
\begin{eqnarray}
\mathrm{Re}\,W^{\mathrm{subcc}} & = &\int_{t_{s}}^{t_{r}}\frac{dt}{\sqrt{z(t)^{2}+x(t)^{2}}}\nonumber\\
 & = &i\int_{t_{i}}^{0}\frac{d\tau}{\sqrt{\tilde{z}(\tau)^{2}+\tilde{x}(\tau)^{2}}}.\label{eq:Re_W_subcc_components}
\end{eqnarray}
For a monochromatic laser pulse (\ref{eq:mono_pulse}), the zeroth-order sub-barrier trajectory ($\ref{eq:complex_traj})$ as a function of $t=t_r+ i \tau$ can be cast
into [see (\ref{eq:aandb}), (\ref{eq:A_and_B})]
\begin{equation}
\begin{split}\tilde{z}(\tau) & =\frac{A_{0}}{\omega}\cos\left(\omega t_{r}\right)\left(\cosh\omega\tau-\cosh\omega t_{i}\right)\\
 &\qquad  +i\left(p_{z}\tau-\frac{A_{0}}{\omega}\sin\omega t_{r}\sinh\omega\tau\right)
\end{split}
\label{eq:z_tau_sub_mono}
\end{equation}
and
\begin{equation}
\tilde{x}(\tau)=ip_{x}\tau.\label{eq:x_tau_sub_mono}
\end{equation}
Substituting $t_{r}^{(1\pm)}$ and $p_{z}^{(1\pm)}$ into $\tilde{z}(\tau)$
(\ref{eq:z_tau_sub_mono}), we find with the help of the conditions (\ref{eq:perpendicular_conditions}) $\tilde{z}^{(1+)}(\tau)=-\tilde{z}^{(1-)}(\tau)$. 
For $\tilde{x}(\tau)$ (\ref{eq:x_tau_sub_mono}) clearly follows
$\tilde{x}^{(1+)}(\tau)=\tilde{x}^{(1-)}(\tau)$. As a consequence, the integrand in 
(\ref{eq:Re_W_subcc_components}) is identical for both trajectories. As the integration starts for both trajectories at $t_i^{(1+)}=t_i^{(1-)}=t_i$ we obtain  $\mathrm{Re}\, W_{\mathrm{c}\perp}^{\mathrm{subcc}(1+)}=\mathrm{Re}\, W_{\mathrm{c}\perp}^{\mathrm{subcc}(1-)}$, i.e.,
the sub-CC phase difference vanishes, and (\ref{eq:phi_c_subcc_perpendicular})
is proved.

In passing, we would like to mention that caution  has to be exercised regarding which saddle-points are chosen for $p_z=0$ if zeroth-order, {\em plain} SFA were used, for which the degeneracy of the saddle points is {\em not} lifted. If the ``usual'' long and short trajectories $(1+)$ and $(2+)$ with $t_{r}^{(1+)}$
and $t_{r}^{(2+)}$ shown in Fig.~\ref{fig:demo}(b) are chosen, $\tilde{z}^{(2+)}(\tau)$ can be expressed by variables of $(1+)$
as
\begin{equation}
\begin{split}\tilde{z}^{(2+)}(\tau) & =-\frac{A_{0}}{\omega}\cos(\omega t_{r}^{(1+)})\left(\cosh\omega \tau-\cosh\omega t_{i}\right)\\
 & \qquad +i\left(p_{z}^{(1+)}\tau-\frac{A_{0}}{\omega}\sin\omega t_{r}^{(1+)}\sinh\omega \tau\right).
\end{split} \label{eq:z2plus2}
\end{equation}
Compared to (\ref{eq:z_tau_sub_mono}), this expression differs by a sign only in the first term. Since $\tilde{x}^{(1+)}(\tau)=\tilde{x}^{(2+)}(\tau)$,
the imaginary parts of $z(t)^{2}+x(t)^{2}$ in the denominator of
(\ref{eq:Re_W_subcc_components}) for $(1+)$ and $(2+)$ are opposite
in signs. It seems a paradox occurs: in the exact perpendicular
direction, integrands are identical since the imaginary part in  (\ref{eq:z2plus2}) is zero,
and thus no phase difference exists. But a  tiny longitudinal momentum
$p_{z}$ would lead to the abrupt appearance of a finite phase
difference, which is unphysical. The problem does not arise in the above employed sub-CC, thanks to the Coulomb distortion of the trajectories in real-space, which enforces $p_{0z}^{(1+)}=-p_{0z}^{(1-)}$, thus determining the relevant saddle-point times to be $(1+)$ and $(1-)$, {\em not} $(1+)$ and $(2+)$.

\subsubsection{Phase difference from sub-barrier propagation without Coulomb interaction:
$\phi_{\mathrm{c}\perp}^{\mathrm{sub},0}$}\label{sub:perp_sub_0}
After having identified the relevant saddle-points $(1+)$ and $(1-)$ we now show that 
\begin{equation}
\phi_{\mathrm{c}\perp}^{\mathrm{sub},0}=\mathrm{Re}\left[W^{\mathrm{sub},0(1+)}-W^{\mathrm{sub},0(1-)}\right]=0.\label{eq:phi_c_sub0_perpendicular}
\end{equation}
Substituting  a monochromatic laser field (\ref{eq:mono_pulse})
 into (\ref{eq:w_sub_0}) yields
\begin{equation}
\begin{split}W^{\mathrm{sub},0} & =i\left(I_{p}+\frac{1}{2}p^{2}+U_{p}\right)t_{i}\\
 & \qquad -\frac{A_{0}p_{z}}{\omega}\left(\cos\omega t_{r}-\cos\omega t_{s}\right)\\
 & \qquad +\frac{A_{0}^{2}}{8\omega}\left(\sin2\omega t_{r}-\sin2\omega t_{s}\right),
\end{split}
\end{equation}
where $p=\sqrt{p_{z}^{2}+p_{x}^{2}}$. The real part reads
\begin{equation}
\begin{split}\mathrm{Re}\, W^{\mathrm{sub},0}= & -\frac{A_0p_z}{\omega}\cos\omega t_r (1-\cosh\omega t_i) \\
 & +\frac{A_0^2}{8\omega} \sin 2\omega t_r (1-\cosh 2\omega t_i)
.
\end{split}
\label{eq:Re_W_sub_0}
\end{equation}
Inserting $p_{z}^{(1\pm)}$
and $t_{r}^{(1\pm)}$, which satisfy  (\ref{eq:perpendicular_conditions}), we find 
\begin{equation}
\mathrm{Re}\,W_{\mathrm{c}\perp}^{\mathrm{sub},0(1-)}=\mathrm{Re}\,W_{\mathrm{c}\perp}^{\mathrm{sub,0(1+)}}, \end{equation}
confirming (\ref{eq:phi_c_sub0_perpendicular}).

\subsubsection{Phase difference from real propagation: $\phi_{\mathrm{c}\perp}^{\mathrm{re}}$} \label{sub:perp_re}
The interference pattern in the perpendicular
direction is entirely determined by the real-time trajectories and the corresponding phase difference
\[
\phi_{\mathrm{c}}^{\mathrm{re}}=W^{\mathrm{re}(1+)}-W^{\mathrm{re}(1-)}.
\]
 Given  real trajectories  $z(t)$, $x(t)$, $v_{z}(t)$, $v_{x}(t)$ for $t\in[t_{r},T_p]$, the action due
to the propagation in real time is
\[
\begin{split}W^{\mathrm{re}}= & -\int_{t_{r}}^{T_{p}}\biggl[\frac{1}{2}\left[v_{z}^2(t)+v_{x}^2(t)\right]\\
 & -\frac{Z}{\sqrt{z^2(t)+x^2(t)}}+I_{p}\biggr]dt.
\end{split}
\]
The symmetry properties for two interfering
trajectories in the perpendicular direction are
\begin{eqnarray}
\begin{aligned}v_{z\perp}^{(1-)}(t)& =-v_{z\perp}^{(1+)}\left(t+\frac{\pi}{\omega}\right),\\ 
v_{x\perp}^{(1-)}\left(t\right)&= v_{x\perp}^{(1+)}\left(t+\frac{\pi}{\omega}\right),\\
z_{\perp}^{(1-)}(t) &=-z_{\perp}^{(1+)}\left(t+\frac{\pi}{\omega}\right),\\
 x_{\perp}^{(1-)}\left(t\right) &=x_{\perp}^{(1+)}\left(t+\frac{\pi}{\omega}\right).
\end{aligned}
\end{eqnarray}
Moreover, $t_{r}^{(1-)}=t_{r}^{(1+)}-{\pi}/{\omega}$. Casting $W_{\perp}^{\mathrm{re}(1-)}$ in terms of the variables of $(1+)$ 
one finds
\begin{eqnarray}
\begin{split}W_{\perp}^{\mathrm{re}(1-)} & =-\int_{t_{r}^{(1+)}}^{T_{p}+\pi/\omega}\biggl\{\frac{1}{2}\left[v_{z\perp}^{(1+)}(t)^2+v_{x\perp}^{(1+)}(t)^2\right]\\
 & \quad  -\frac{Z}{\sqrt{{z_{\perp}^{(1+)}}(t)^2+{x_{\perp}^{(1+)}}(t)^2}}+I_{p}\biggr\}\, dt \\
& = - \int_{t_{r}^{(1+)}}^{T_{p}+\pi/\omega} I(t)\,d t.
\end{split}
\end{eqnarray}
The phase difference $\phi_{\mathrm{c}\perp}^{\mathrm{re}}=W_{\perp}^{\mathrm{re}(1+)}-W_{\perp}^{\mathrm{re}(1-)}$
is thus given by
\begin{equation}
\phi_{\mathrm{c}\perp}^{\mathrm{re}}=\left(-\int_{t_{r}^{(1+)}}^{T_{p}}+\int_{t_{r}^{(1+)}}^{T_{p}+\pi/\omega}\right) I(t)\, d t= \int_{T_{p}}^{T_{p}+\pi/\omega} I(t)\, dt,
\end{equation}
independent of the specific saddle-point pair under consideration. 
As $T_{p}\rightarrow\infty$ the asymptotic position in polarization direction remains finite, $z(t)|_{t\rightarrow\infty}\rightarrow z_{\infty}$, but $x(t)|_{t\rightarrow\infty}\rightarrow\infty$, leading
to a vanishing phase difference due to the Coulomb potential. In the kinetic energy term of the integrand,
$v_{z}(t)|_{t\rightarrow\infty}\rightarrow A(t)$ and $v_{x}(t)|_{t\rightarrow\infty}\rightarrow p_{x}$,
so that
\begin{equation}
\begin{aligned}\phi_{\mathrm{c}\perp}^{\mathrm{re}} & =\int_{T_{p}}^{T_{p}+\pi/\omega}\left(\frac{1}{2}p_{x}^{2}+\frac{1}{2} A^2(t) + I_{p}\right)dt\\
 & =\frac{\pi}{\omega}\left(\frac{1}{2}p_{x}^{2}+I_{p}\right) + \frac{1}{2}\int_{T_{p}}^{T_{p}+\pi/\omega} A^2(t)\, dt .
\end{aligned}
\end{equation}
Using $p_{x}=\sqrt{2\left(n\omega-U_{p}-I_{p}\right)}$ from the energy conservation for ATI in the perpendicular direction with $n$ the number of absorbed photons 
and $U_{p}=\frac{1}{2}\frac{1}{2\pi/\omega}\int_{0}^{2\pi/\omega}A^2(t)\,dt$
we find
\begin{equation}
\phi_{\mathrm{c}\perp}^{\mathrm{re}}=n\pi.\label{eq:phi_c_re_perpendicular}
\end{equation}
This result  agrees with the findings in Ref.~\cite{ref_PRL_perpendicular_carpet} that every
other ATI ring is destructively ``interfered away'' in perpendicular direction. Because of the pre-factor $C_{\boldsymbol{p}}(t_{s}^{(\alpha)})$ in (\ref{spsum})  the interference is destructive for odd $n$ in the case of   even-parity initial states and {\em vice versa}.

The interference concerning the ATI rings in perpendicular direction predicted by (\ref{eq:phi_c_re_perpendicular}) is confirmed by the results  for ionization of H(1s) in a  9-cycle
trapezoidal-envelope laser pulse of 800\,nm wavelength  shown in  Fig.~\ref{fig:spec_2D_const_800nm_1e14}.  The other laser parameters are given in the figure caption.

\section{ATI on-ring structure}\label{sec:ationring}
\begin{figure}
\includegraphics[scale=0.6]{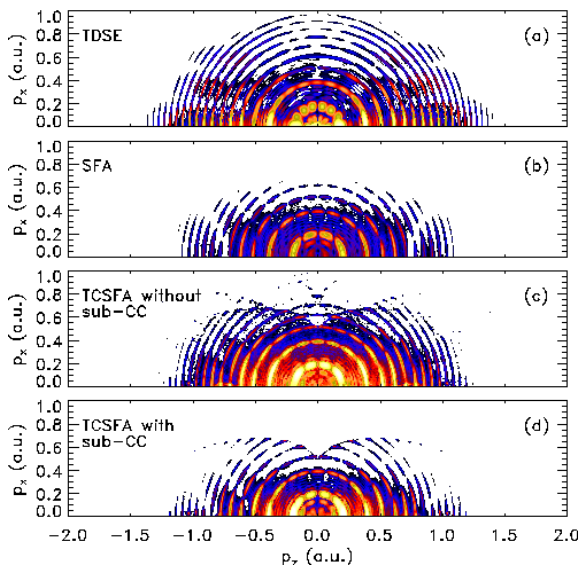}

\caption{(color online). Same as in Fig.~\ref{fig:spec_2D_sin2_2um_1e14_nc4} but for a 9-cycle,
trapezoidal-envelope laser (2, 5, and 2 cycles for up-ramping, constant-envelope,
and down-ramping, respectively). The wavelength is $\lambda=800$\,nm, the intensity is $10^{14}$ W/cm$^{2}$. In the spectrum
calculated using the TCSFA with sub-CC (d) only trajectories of type T1 and T2 in the nomenclature of \cite{ref_PRL_TCSFA_LES} have been considered \cite{remark:onlyT1T2}. \label{fig:spec_2D_const_800nm_1e14}}
\end{figure}

ATI peaks appear as rings of radius 
\begin{equation}
p=\sqrt{2\left(n\omega-U_{p}-I_{p}\right)}
\end{equation} 
in the photoelectron momentum spectra. 
It is known from experiments that the ionization probability
along the ATI rings is not isotropic but exhibits nodal structures. These structures, also visible in Fig.~\ref{fig:spec_2D_const_800nm_1e14}, have been analyzed recently within the SFA \cite{ref_PRL_perpendicular_carpet}. 
 The radius of the ATI rings is well reproduced already by the plain SFA, which is verified by comparing Fig.~\ref{fig:spec_2D_const_800nm_1e14}(b) with the \emph{ab initio} TDSE result Fig.~\ref{fig:spec_2D_const_800nm_1e14}(a).
 However, the positions of the nodes along low-order ATI rings  do not agree well. Not even the {\em number} of minima and maxima agrees. Take, e.g.,  the second ATI ring ($p=0.3939$, absorption of 14 photons). The TDSE result in (a) shows 9 maxima along that ring, the plain SFA result in (b) only 7. The real-time Coulomb-corrected SFA in (c) does not affect the number of maxima. Instead, the sub-CC SFA in (d) does give the correct number. Along the next ATI-ring (absorption of 15 photons) TDSE and plain SFA agree already (10 maxima), and TCSFA with or without sub-CC does not change the nodal structure. Along the first ATI ring the TDSE shows 8 maxima, which is not properly reproduced, even with the sub-CC TCSFA, probably because our sub-barrier Coulomb correction involving only the zeroth-order plain-SFA trajectories is insufficient.

\begin{figure}
\includegraphics[width=0.35\textwidth]{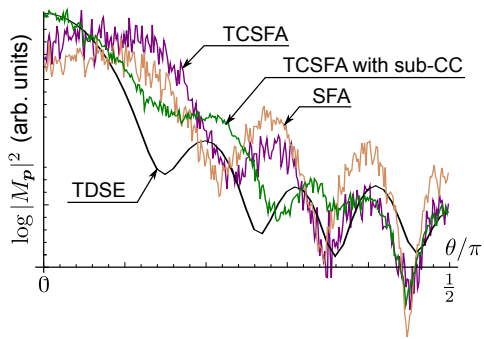}

\caption{(color online). The ionization probability along the second ATI ring 
 with $p=0.3939$  in Fig.~\ref{fig:spec_2D_const_800nm_1e14} according to the  \emph{ab initio} TDSE, the (plain) SFA, the TCSFA (without sub-CC), and the TCSFA with sub-cc result. The polar angle $\theta$ is in the
range $[0,\pi/2]$, $\theta=0$ corresponding to $p_x=0$, and $\theta=\pi/2$  corresponding to $p_z=0$.  The plain SFA result is as noisy as the TCSFA because it  was calculated using the same statistical sampling approach as for the TCSFA (which is explained briefly in Sec.~\ref{sec:repart} and in more detail in \cite{ref_PRL_TCSFA_LES}).  \label{fig:on_ring_nodal}}
\end{figure}

Figure~\ref{fig:on_ring_nodal} shows the momentum distribution
along the second ATI ring 
for the \emph{ab initio} TDSE, the SFA, and the TCSFA with or without sub-CC.
TCSFA with sub-CC agrees  well with the \emph{ab initio} TDSE result.
The results from the TCSFA without sub-CC and the plain SFA are similar. They agree well with the TDSE around  $\theta=\pi/2$ but deviate for smaller angles. As discussed above, there is one maximum less in the plain SFA and the TCSFA without sub-CC, as compared to the TDSE and the TCSFA with sub-CC. Hence we conclude that the nodal structure of low-order ATI rings is affected by the sub-CC. Higher-order ATI rings may be effected by the side-lobes (clearly visible in Fig.~\ref{fig:spec_2D_const_800nm_1e14}(a) and discussed in  \cite{ref_huisman_science}).

\section{Summary}\label{sec:summ}
We extended the previously introduced trajectory-based Coulomb-corrected strong field approximation towards the inclusion of sub-barrier Coulomb effects. To that end we used the plain strong field approximation quantum orbits and inserted them into the Coulomb-corrected sub-barrier action. The (additional) phase-difference between the so-called ``long'' and ``short'' trajectories introduced in that way has a pronounced qualitative effect on the photoelectron spectra, namely a phase shift of  $Z\pi/\sqrt{2I_{p}}$ along the polarization axis for linear polarization in the long-wavelength limit. In the case of ionization from the ground state of atomic hydrogen or hydrogen-like ions, the phase shift is $\pi$, turning constructive intra-cycle interference maxima into destructive interference minima and {\em vice versa}. We benchmarked the predictions of the trajectory-based Coulomb-corrected strong field approximation including sub-barrier effects by comparisons with {\em ab initio} solutions of the  time-dependent Schr\"odinger equation.
We also showed that the sub-barrier Coulomb correction affects the nodal structure along low-order above-threshold ionization rings. 

Our semi-classical method is applicable to sufficiently long wavelengths ($I_p/\omega\gg 1$). However, the nuclear charge $Z$ (and thus $I_p$) has to be small enough to render the zeroth-order orbit a good approximation for the sub-barrier Coulomb correction. Moreover, the  trajectory-based Coulomb-corrected strong field approximation        displays too many unphysical caustic structures in the spectra if the Keldysh parameter $\sqrt{I_p/(2U_p)}$ is not small enough (i.e., around $1$ or smaller). It thus turns out that our method is applicable in a regime where the direct {\em ab initio} solution of the  time-dependent Schr\"odinger equation in position space is stretched to its limits: long wavelengths and high ponderomotive energy.

Our method is, in principle, also applicable to more complex systems for which strong field approximations already have been or can be developed. The most straightforward extension is towards two Coulomb-centers (e.g., H$_2^+$). Also non-sequential ionization in effective two-electron systems is accessible. However, due to the higher-dimensional momentum space to be sampled in such systems it will be numerically very demanding to obtain differential photoelectron spectra with good enough statistics. The influence of many-electron polarization effects could be investigated by incorporating a polarization potential into the equations of motion for the quantum trajectories \cite{pfeiffer,maurer}.


\begin{acknowledgments}
We thank S.V.\ Popruzhenko for fruitful discussions. The work was supported by the Deutsche Forschungsgemeinschaft
(SFB 652). T.-M. Yan
acknowledges support from the International Max Planck Research School
for Quantum Dynamics (IMPRS-QD) in Heidelberg.\end{acknowledgments}


\begin{thebibliography}{References}
\bibitem{realorimaginary} In other approaches this is not necessarily the case (see, e.g., \cite{olga}).
\bibitem{linuxcluster} On 100 nodes of a present-day linux cluster our calculations took two hours.
\bibitem{remark:sidelobes}The origin of the almost horizontally aligned side lobes, most clearly visible for $p_z>0$, are discussed in \cite{ref_huisman_science}.
\bibitem{remark:prefactorphases}As the  prefactors $C_{\boldsymbol{p}}(t_{s}^{(\alpha)})$ involve the initial state, extra phase differences may occur when starting from excited states.
\bibitem{remark:realpartunimportant} We have checked numerically that Eq.~(\ref{eq:approxp1p2p}) is valid along the polarization axis in the long-wavelength regime.
\bibitem{remark:onlyT1T2} Exaggerated caustic structures appear that spoil the interference structure if T3 and T4 trajectories  (in the nomenclature of \cite{ref_PRL_TCSFA_LES}) are included. The situation improves with increasing $I_p/\omega$ and decreasing Keldysh parameter $\sqrt{I_p/(2U_p)}$.
\end{thebibliography}
\end{document}